\documentclass[prl,aps,showpacs,twocolumn,floatfix]{revtex4}
\usepackage{epsfig} \usepackage{graphics} \usepackage{bm}
\usepackage{amssymb}
\usepackage{graphicx}
\begin{document}

\preprint{Lebed-Rapids-LN}

\title{Can a Magnetic Field Destroy a Spin-Density-Wave Phase in a
Quasi-One-Dimensional Conductor?}

\author{A.G. Lebed$^*$}

\affiliation{Department of Physics, University of Arizona, 1118 E.
4-th Street, Tucson, AZ 85721, USA}

\begin{abstract}
It is known that, in a pure one-dimensional case,
Charge-Density-Wave (CDW) phase is destroyed by a magnetic field,
whereas Spin-Density-Wave (SDW) one does not feel the field. In
reality, SDW phase is often observed in quasi-one-dimensional
(Q1D) conductors due to the so-called "nesting" property of their
electron spectra. We show that, in the latter case, a high
magnetic field generates some "anti-nesting" term in a Q1D
electron spectrum, which destroys SDW phase. We suggest to perform
the corresponding experiments in SDW phases of the real Q1D
organic conductors with chemical formula (TMTSF)$_2$X (X=PF$_6$,
ClO$_4$, etc.).

\end{abstract}


\maketitle

We recall that a pure 1D metal is unstable with respect to the
so-called Peierls transition into some Density-Wave (DW) state
[1,2]. The DW state can be characterized by either a spatial
modulation of a charge [i.e., Charge-Density-Wave (CDW) phase] or
spatial modulation of a spin [i.e., Spin-Density-Wave (SDW) one]
[2-4]. Let us discuss in a brief the above mentioned phenomenon
[1,2]. Indeed, near two plane Fermi surfaces (FS), it is possible
to linearize 1D spectrum,
\begin{equation}
\epsilon(p_x) = -2t_a \cos(p_xa^*),
\end{equation}
in the following way:
\begin{equation}
\epsilon^{\pm}(p_x) = \pm v_F (p_x \mp p_F) ,
\end{equation}
where $v_F$ and $p_F$ are the Fermi velocity and momentum,
correspondingly, $a^*$ is a lattice constant. (Note that here and
everywhere below we make use for actual electron spectrum its
tight-binding model, since we apply our results to organic
conductors [3,4], where this model is known to work well [3]).

It is important that the electron spectrum (2) possesses the
following special (i.e., "nesting") property of electron-hole
pairing,
\begin{equation}
\epsilon^{+}(\Delta p_x) + \epsilon^{-}(\Delta p_x) =0,
\end{equation}
which makes some DW with wave vector $2p_F$ to be a ground state
at low enough temperatures [1,2]. In an external magnetic field,
the electron spectra (1) and (2) split into two branches due to
the Pauli spin-splitting effect:
\begin{equation}
\epsilon_{\sigma}^{\pm}(p_x) = \pm v_F (p_x \mp p_F) - \sigma
\mu_B H ,
\end{equation}
where $\sigma = \pm 1$ for spin up(down), $\mu_B$ is the Bohr
magneton. From Eq.(4), it directly follows that the condition (3)
for electron-hole pairing in a magnetic field is not changed for
SDW phase and changed for CDW one. Therefore, we can make the
well-known conclusion that SDW phase is stable in the presence of
the Pauli spin-splitting effects in a magnetic field [4,5],
whereas CDW one is destroyed by the field [6].

Let us consider a quasi-one-dimensional (Q1D) conductor with the
folloiwng electron spectrum [3],
\begin{equation}
\epsilon({\bf p}) = -2t_a \cos(p_xa^*/2)-2t_b \cos(p_y b^*)-2t_c
\cos(p_z c^*) ,
\end{equation}
where $t_a \gg t_b \gg t_c$. Near two open sheets of the FS, it
can be linearized as
\begin{equation}
\epsilon^{\pm}({\bf p}) = \pm v_F( p_x \mp p_F) -2t_b \cos(p_y
b^*)-2t_c \cos(p_z c^*) .
\end{equation}
It is important that electron spectrum (6) still possesses the
above discussed "nesting" electron-hole symmetry, since for Eq.(6)
the following equation is valid:
\begin{equation}
\epsilon^{+}(\Delta p_x,p_y,p_z) + \epsilon^{-}(\Delta
p_x,p_y+\pi/b^*,p_z+\pi/c^*) = 0 .
\end{equation}
It is possible to make sure [7] that the "nesting" property (7)
corresponds to a stability of some DW with the wave vector:
\begin{equation}
Q^0 = (2p_x,\pi/b^*,\pi/c^*).
\end{equation}
As also suggested in Ref.[7], the "nesting" property (7) is
responsible for the appearance of SDW in the real Q1D conductors
from chemical family (TMTSF)$_2$X, where X=PF$_6$, ClO$_4$, etc.
[3,4].

In Refs.[5,7], a more realistic electron spectrum is considered:
\begin{eqnarray}
\epsilon^{\pm}({\bf p}) = \pm v_F( p_x \mp p_F) -2t_b \cos(p_y
b^*)
\nonumber\\
-2t'_b \cos(2p_y b^*) -2t_c \cos(p_z c^*),
\end{eqnarray}
where it includes also the next-neighbor electron jumping in
tight-binding model, $t'_b \ll t_b$. The electron spectrum (9)
contains the so-called "anti-nesting" term, $2t'_b \cos(2p_y
b^*)$. This term destroys the ideal "nesting" condition (7) and,
thus, at large enough values of the parameter $t'_b$, restores a
metallic phase. For the theory of experimentally observed in the
Q1D conductors (TMTSF)$_2$X  the Field-Induced Spin-Density-Wave
(FISDW) phases [8,9], where the "anti-nesting" term in Eq.(9)
plays a central role, see Refs.[5,10]. For further development, it
is important that electron spectra (6),(9) still show the same
properties in a parallel to the conducting chains magnetic field,
where the orbital effect [5] is negligible. More specifically, SDW
phase still does not fill the Pauli spin-splitting effects,
whereas CDW one is destroyed by them.

The goal of our Rapid Communication is to consider unexpected
novel effect - a destruction of SDW phase in Q1D conductors with
"nesting" properties (7) by the Pauli spin-splitting effect. Here,
we restrict our calculations by case of a parallel magnetic field
to avoid complications due to possible appearance of the FISDW
phases as a result of orbital electron quantization [5,8-10]. Note
that below we consider model, which can be solved analytically,
and suggest to perform the corresponding experiments in the Q1D
organic conductors (TMTSF)$_2$X. The physical meaning of the
suggested phenomenon is as follows. We show that, due to non-zero
$Q^0_y$ component of SDW wave vector (8) and due to non-linearity
of electron spectrum along the conducting chains, the Pauli
spin-splitting effect generates a special "anti-nesting" term.
This term increases with a growing magnetic field and eventually
destroys SDW phase. We stress that the above mentioned statement
is against a common belief that the Pauli spin-splitting effect
does not influence SDW phase.

Below, we consider the following 2D model of Q1D spectrum in the
(TMTSF)$_2$X conductors in a parallel magnetic field,
\begin{eqnarray}
\epsilon_{\sigma}({\bf p}) = -2t_a \cos(p_xa^*/2)-2t_b \cos(p_y
b^*)
\nonumber\\
 -2t'_b \cos(2 p_y b^*) - \mu_B \sigma H .
\end{eqnarray}
[Note that, as well known [3,4,5,7,10], 2D model (10) well
describes the SDW and FISDW phases in these conductors, since $t_b
\gg t_c$ in Eq.(5)]. In contrast to the all existing works, we do
not linearize the electron spectrum along the conducting ${\bf
a^*}$ axis near two sheets of the FS, but also take into account
the next quadratic term:
\begin{eqnarray}
\epsilon^+(p_x) = v_F (p_x-p_F) + \alpha (p_x-p_F)^2 ,
\nonumber\\
\epsilon^-(p_x) = -v_F (p_x+p_F) + \alpha (p_x+p_F)^2 ,
\end{eqnarray}
where
\begin{equation}
v_F = \frac{t_a a^*}{\sqrt{2}} , \ \ \ \alpha = \frac{t_a
(a^*)^2}{4 \sqrt{2}}.
\end{equation}
[In Eqs.(10)-(12), we take into account that $p_F = \pi/2a^*$ in
the (TMTSF)$_2$X conductors.]

Let us derive electron energy spectra in a parallel magnetic field
near two sheets of the FS by means of Eqs.(10)-(12). To this end,
first let us rewrite Eq.(10) in the following way:
\begin{eqnarray}
&&\epsilon^+_{\sigma}({\bf p}) = v_F (p_x-p_F) + \alpha
(p_x-p_F)^2
\nonumber\\
&&-2t_b \cos(p_y b^*) -2t'_b \cos(2 p_y b^*) - \mu_B \sigma H
\end{eqnarray}
and
\begin{eqnarray}
&&\epsilon^-_{\sigma}({\bf p}) = -v_F (p_x+p_F) + \alpha
(p_x+p_F)^2
\nonumber\\
&&-2t_b \cos(p_y b^*) -2t'_b \cos(2 p_y b^*) - \mu_B \sigma H .
\end{eqnarray}
Then, we define the shapes of two sheets of the FS for the value
of small parameter $\alpha =0$ in Eqs.(13) and (14) (i.e., in the
linear approximation):
\begin{eqnarray}
(p_x-p_F) = \frac{2t_b \cos(p_y b^*) +2t'_b \cos(2 p_y b^*) +
\mu_B \sigma H}{v_F}
\end{eqnarray}
and
\begin{eqnarray}
(p_x+p_F) = - \frac{2t_b \cos(p_y b^*) +2t'_b \cos(2 p_y b^*) +
\mu_B \sigma H}{v_F} .
\end{eqnarray}
Now, let us put the obtained values of $p_x-p_F$ and $p_x+p_F$,
given by Eqs.(15) and (16), only in terms, which contain the small
parameter, $\alpha \neq 0$, in Eqs.(13) and (14). As a result, for
$t_b, \mu_B H \gg t'_b$, we obtain the following electron spectra
near two sheets of the FS in the quadratic approximation:
\begin{eqnarray}
\epsilon^+_{\sigma}({\bf p}) = v_F(p_x-p_F) + t^+_{b}(p_y,\sigma)
-\mu_B \sigma H + \Delta \epsilon ,
\nonumber\\
t^+_b(p_y, \sigma) = - 2t_b \cos(p_y b^*) + 2\tilde t'_b \cos(2
p_y b^*)
\nonumber\\
+ 2t_H \sigma \cos(p_y b^*)
\end{eqnarray}
and
\begin{eqnarray}
\epsilon^-_{\sigma}({\bf p}) = - v_F(p_x+p_F) +
t^-_{b}(p_y,\sigma) -\mu_B \sigma H + \Delta \epsilon ,
\nonumber\\
t^-_{b}(p_y,\sigma) = - 2t_b \cos(p_y b^*)+ 2 \tilde t'_b \cos(2
p_y b^*) \nonumber\\
 + 2t_H \sigma \cos(p_y b^*)  ,
\end{eqnarray}
where $ t_H = \mu_B H t_b/(\sqrt{2} t_a), \ \tilde t'_b = - t'_b +
t^2_b/(2 \sqrt{2}t_a), \ \Delta \epsilon = \mu^2_B H^2/(2
\sqrt{2}t_a)$. Note that Eqs.(17) and (18) contain magnetic field
dependent term, $t_H \sim H$, which, for SDW pairing, breaks the
electron-hole pairing condition (7) and, thus, destroys SDW phase
at high magnetic fields. In contrast, terms $-\mu_BH$ and $\Delta
\epsilon$ in Eqs.(17) and (18) do not destroy SDW pairing. Indeed,
term $-\mu_BH$ disappears for SDW pairing, whereas term $\Delta
\epsilon$ just shifts the wave vector of SDW phase.

Our goal is to describe quantitatively the destruction of SDW by a
magnetic field due to "anti-nesting" term in Eqs.(17) and (18),
which contains magnetic field dependent parameter $t_H$. Let us
calculate the linear response of our system to the following
external field, corresponding to SDW pairing.
\begin{equation}
\hat h ({\bf Q}) = (\hat \sigma_x)_{\alpha \beta} \exp(i {\bf Q}
{\bf r}) \ ,
\end{equation}
We do this in a similar way, as it is done in Ref.[5] for
different Q1D spectrum without the above mentioned magnetic field
dependent term. In mean field approximation, we obtain for
susceptibility the so-called Stoner's equation:
\begin{equation}
\chi ({\bf Q}) = \frac{\chi_0 ({\bf Q})}{[1 - g \chi_0 ({\bf
Q})]}.
\end{equation}
Here $g$ is the effective electron coupling constant, $\chi_0
({\bf Q})$ is susceptibility of non-interacting electrons:
\begin{eqnarray}
&&\chi_0({\bf Q}) = T \sum_{\omega_m} \sum_{\sigma} \int
\frac{dp_y}{2 \pi} \int dx_1 g^{++}(i \omega_n, p_y; x, x_1;
\sigma)
\nonumber\\
&&\times g^{--}(i \omega_n, p_y -Q_y; x_1, x; -\sigma) ,
\end{eqnarray}
where $\omega_n$ is the so-called Matsubara's frequency [11].

In Eq.(21), slow varying parts of the electron Green's functions
near two sheets of Q1D FS are related to the electron Green's
functions by the following equation:
\begin{eqnarray}
G^{++}(i \omega_n, p_y; x, x_1; \sigma) = e^{ip_F(x-x_1)}
\nonumber\\
\times g^{++}(i \omega_n, p_y; x, x_1; \sigma),
 \end{eqnarray}
\begin{eqnarray}
 G^{--}(i \omega_n, p_y; x, x_1; \sigma) = e^{-ip_F(x-x_1)}
 \nonumber\\
\times g^{--}(i \omega_n, p_y; x, x_1; \sigma).
\end{eqnarray}
Slow varying parts of Green's functions of non-interacting
electrons are possible to determine by using the method similar to
that suggested in Ref.[5]. As a result, we obtain the following
equations:
\begin{eqnarray}
\biggl[i \omega_n +i v_F \frac{d}{dx} -t^+_b (p_y, \sigma) +\mu_B
\sigma H - \Delta \epsilon \biggl]
\nonumber\\
\times g^{++}(i \omega_n, p_y; x, x_1; \sigma) = \delta (x-x_1) \
,
\end{eqnarray}
\begin{eqnarray}
\biggl[i \omega_n - i v_F \frac{d}{dx} -t^-_b (p_y, \sigma) +\mu_B
\sigma H - \Delta \epsilon \biggl]
\nonumber\\
\times g^{--}(i \omega_n, p_y; x, x_1; \sigma) = \delta (x-x_1) \
,
\end{eqnarray}
where $\delta(x-x_1)$ is the Dirac's delta-function. It is
important that Eqs.(24) and (25) can be exactly solved:
\begin{eqnarray}
g^{++}(i \omega_n, p_y; x, x_1; \sigma)= \frac{sgn (\omega_n)}{i
v_F} \exp \biggl[ -\frac{\omega_n(x-x_1)}{v_F}
\nonumber\\
-\frac{i}{v_F} t^+_b (p_y, \sigma)(x-x_1) +\frac{i}{v_F}\mu_B
\sigma H(x-x_1)
\nonumber\\
-\frac{i}{v_F} \Delta \epsilon (x-x_1) \biggl] , \ \ \omega_n
(x-x_1) > 0,
\end{eqnarray}
\begin{eqnarray}
g^{--}(i \omega_n, p_y; x, x_1; \sigma)= \frac{sgn (\omega_n)}{i
v_F} \exp \biggl[ \frac{\omega_n(x-x_1)}{v_F}
\nonumber\\
+\frac{i}{v_F} t^-_b (p_y, \sigma)(x-x_1) -\frac{i}{v_F}\mu_B
\sigma H(x-x_1)
\nonumber\\
+\frac{i}{v_F} \Delta \epsilon (x-x_1) \biggl] , \ \ \omega_n
(x-x_1) < 0 .
\end{eqnarray}

Now, let us substitute the known Green's functions [i.e., Eqs.
(26) and (27)] into Eqs.(20) and (21). After straightforward but
rather lengthy calculations, we obtain the following equation,
which determines a stability region of SDW phase:
\begin{eqnarray}
\frac{1}{g}= \max_{\tilde k, \Delta t} \int^{\infty}_d \frac{2 \pi
T_c dz}{v_F \sinh \biggl( \frac{2 \pi T_c z}{v_F} \biggl)} \biggl<
\cos \biggl[ \frac{4 \Delta t}{v_F} \sin(p_y b^*) \ z
\nonumber\\
- \frac{4 \tilde t'_b}{v_F} \cos(2p_yb^*) \ z + \tilde k z \biggl]
\cos \biggl[\frac{4t_H}{v_F} \cos(p_y b^*) \ z \biggl]
\biggl>_{p_y} ,
\end{eqnarray}
where $Q_y = \pi/b^* + q \ (qb^* \ll 1)$, $\Delta t = t_b q
b^*/2$, $\tilde k = k - 2 \Delta \epsilon /v_F$, $d$ is a cut-off
distance; $<...>_{p_y}$ stands for averaging procedure over
variable $p_y$. Note that, in Eq.(28), we maximize SDW transition
temperature, $T_c$, with respect to longitudinal, $k$, and
transverse, $Q_y$, wave vectors under condition that $t_b \gg
t'_b$.

As follows from Eq.(28), the last term with $t_H$ will eventually
destroy SDW phase at high magnetic fields. In this Rapid
Communication, we do not intent to investigate Eq.(28) for all
possible cases and all possible values of the parameters $t_b$,
$T_c$, and $\tilde t'_b$. Our goal is to demonstrate that high
enough magnetic field indeed destroys SDW phase even at $T_c=0$
and estimate the corresponding critical field. To this end, we
consider the case of very high magnetic fields, where $t_H \gg
\tilde t'_b$, at $T_c=0$. As we show below, this case can be
analytically solved. Indeed, at $t_H \gg \tilde t'_b$ and $T_c=0$,
we have from Eq.(28):

\begin{eqnarray}
\frac{1}{g}= \max_{\tilde k, \Delta t} \int^{\infty}_d
\frac{dz}{z}  \biggl< \cos \biggl[ \frac{4 \Delta t}{v_F} \sin(p_y
b^*)\ z+ \tilde k z \biggl]
\nonumber\\
\times \cos \biggl[\frac{4t_H}{v_F} \cos(p_y b^*) \ z \biggl]
\biggl>_{p_y} .
\end{eqnarray}
For $\tilde t'_b =0$ and $H=0$, from Eq.(28), we can obtain
another simple equation, which connects electron coupling
constant, $g$, with SDW transition temperature at $H=0$, $T_{c0}$:
\begin{equation}
\frac{1}{g}=  \int^{\infty}_{\frac{2 \pi T_{c0}d}{v_F}} \frac{
dz}{\sinh (z)} .
\end{equation}
Our current problem is to find maximum of the integral (29) over
longitudinal and transverse momenta. This maximum defines the
critical field, $H_0$, which can be expressed through $T_{c0}$,
using Eq.(30).

So, let us first consider Eq.(29). By means of simple but rather
lengthy calculations, it is possible to demonstrate that it is
equivalent to the following simpler equation:
\begin{eqnarray}
\frac{1}{g}= \max_{\tilde k, \Delta t} \int^{\infty}_d
\frac{dz}{z} \ J_0 \biggl[ \sqrt{ \biggl( \frac{4 \Delta t}{v_F}
\biggl)^2 + \biggl( \frac{4t_H}{v_F} \biggl)^2 } \  z \biggl]
\nonumber\\
\times \cos(\tilde k z) ,
\end{eqnarray}
where we use the following formula for the zeroth-order Bessel
function [12]:
\begin{equation}
J_0(z) = \int^{\pi}_{-\pi} \frac{d \phi}{2 \pi} \exp(i z \sin
\phi).
\end{equation}
From Eq.(31), it directly follows that the integral (31) takes its
maximum at $\Delta t =0$ (i.e., for transverse component of the
SDW wave vector $Q_y=\pi/b^*$). Therefore, Eq.(31) can be
simplified as
\begin{equation}
\frac{1}{g}= \max_{\tilde k_1} \int^{\infty}_{\frac{4t_Hd}{v_F}}
\frac{dz}{z} \ J_0 (z) \ \cos(\tilde k_1 z) , \ \ \tilde k_1 =
\frac{v_F \tilde k}{4 t_H} .
\end{equation}

Here, we express the inverse electron coupling constant through
the SDW transition temperature in the absence of $\tilde t'_b$ and
magnetic field, $H=0$, and cutoff distance, $d$, using Eq.(30).
Exact integration of integral (30) over variable $z$ gives us the
following relationship in the so-called logarithmic approximation:
\begin{equation}
\frac{1}{g}=  \ln \biggl( \frac{v_F}{\pi T_{c0} d} \biggl) .
\end{equation}

Our task now is to find maximum of Eq.(33) with respect to
variable $\tilde k_1$ and express the critical magnetic field for
destruction of SDW, $H_0$, as a function of $T_{c0}$ by means of
Eq.(34). It is easy to rewrite Eq.(33) for small values of the
cutoff parameter, $d$, in the following way:
\begin{eqnarray}
\frac{1}{g}= \max_{\tilde k_1} \biggl\{ \int^{\infty}_0
\frac{dz}{z} \ J_0(z) \ [\cos(\tilde k_1 z)-1]
\nonumber\\
+ \int^{\infty}_0 \frac{dz}{z} \ [J_0(z)-\cos(\tilde k_1 z)] +
\int^{\infty}_{d \tilde k} \frac{dz}{z} \ \cos(z) \biggl\} .
\end{eqnarray}
To simplify (35), we use the following mathematical formulas [12]:
\begin{equation}
\int^{\infty}_0 \frac{dz}{z} \ [J_0(z) - \cos(\alpha z)]= \ln(2
\alpha), \ \alpha > 0 ,
\end{equation}
\begin{equation}
\int^{\infty}_0 \frac{dz}{z} \ [1 - \cos(\alpha z)] \ J_0(\beta
z)= arccosh (\frac{\alpha}{\beta}), \ 0 < \beta < \alpha ,
\end{equation}
\begin{equation}
\int^{\infty}_0 \frac{dz}{z} \ [1 - \cos(\alpha z)] \ J_0(\beta
z)= 0, \ 0 < \alpha < \beta,
\end{equation}
\begin{equation}
- \int^{\infty}_x \frac{\cos(z)}{z} \ dz = C + \ln(x) + \int^x_0
\frac{\cos(z)-1}{z} \ dz ,
\end{equation}
where $C =\ln(\gamma)$ is the so-called Euler constant. As
directly follows from Eqs.(36)-(39), the integral (35) has maximal
value at $|\tilde k| < 4t_H/v_F$, which is equal to
\begin{equation}
\frac{1}{g} = \ln \biggl(\frac{v_F}{2 \gamma t_H d} \biggl),
\end{equation}
where $\gamma \approx 1.78$. Comparison of Eq.(34) and (40)
results in the following value of magnetic field, $H_0$, which
destroys SDW phase at $T=0$:
\begin{equation}
t_{H_0} = \frac{\pi T_{c0}}{2 \gamma}, \ \ \  H_0 =
\frac{1}{\mu_B} \biggl( \frac{\pi T_{c0}}{2 \gamma} \biggl)
\biggl(\frac{\sqrt{2}t_a}{t_b} \biggl) \ .
\end{equation}

To summarize, in the Rapid Communication, we have shown for the
first time that magnetic field generates some "anti-nesting" term
in Q1D conductors due to the Pauli spin-splitting effect. This
term destroys SDW phases, which exist in some Q1D conductors due
to the "nesting" condition. We suggest to perform the
corresponding experiments in the organic conductors (TMTSF)$_2$X.
Let us estimate the critical magnetic field, which destroys SDW
phase. From Eq.(41), it follows that at ambient pressure, where
$T_{c0} = 12 \ K$ in the (TMTSF)$_2$PF$_6$, the critical magnetic
field is $H_0 = 185 \ T$. Although such high magnetic field is
experimentally available (see, for example, Refs.[13],[14]), we
recommend to apply pressure to decrease the value of $H_0$.
Indeed, at pressure $P= 5 \ kbar$, the SDW transition temperature
in the (TMTSF)$_2$PF$_6$ conductor becomes $T_{c0} = 5 \ K$ [15]
and, thus, the critical magnetic field can be estimated as $H_0=77
\ T$. Here, let us discuss in a brief the validity of the above
suggested estimations by Eq.(41) of the critical magnetic fields
to destroy SDW phase in real Q1D compound (TMTSF)$_2$PF$_6$. Note
that, in Eq.(41), we don't explicitly take into account the first
"anti-nesting" term, containing unknown parameter $\tilde t'_b$
[see Eq.(13)-(18)]. Our application of Eq.(41) to real compound
(TMTSF)$_2$PF$_6$ is based on the suggestion that both
"anti-nesting" terms independently decrease SDW transition
temperature. This suggestion is based on the fact that the two
"anti-nesting" terms have different momentum dependence, and,
thus, cannot, for example, cancel each other. Of course, this is
just a reasonable suggestion and, therefore, the above mentioned
calculations of the values of the critical magnetic fields in the
(TMTSF)$_2$PF$_6$ at ambient pressure and $P=5 \ kbar$  are just
some reasonable estimations.

 We are  thankful to N.N. Bagmet (Lebed) for useful discussions.

$^*$Also at: L.D. Landau Institute for Theoretical Physics, RAS, 2
Kosygina Street, Moscow 117334, Russia.


\begin{references}

\bibitem{Peierls}
R.E. Peierls, {\it Quantum Theory of Solids} (Oxford University
Press, London, 1955).

\bibitem{Jerome-1}
D. Jerome and H. Shultz, Adv. Phys. \textbf{31}, 299 (1982).

\bibitem{Yamaji-1}
T. Ishiguro, K. Yamaji, and G. Saito, {\it Organic
Superconductors}, 2nd edn. (Springer, Berlin, 1998).

\bibitem{Lebed-1}
{\it The Physics of Organic Superconductors and Conductors},
edited by A.G. Lebed (Springer, Berlin, 2008).


\bibitem{Lebed-2}
L.P. Gor'kov and A.G. Lebed, J. Phys. (Paris) Lett. \textbf{45},
L-433 (1984).


\bibitem{Ditrich-1}
W. Dietrich and P. Fulde, Z. Phys. \textbf{265}, 239 (1973).


\bibitem{Yamaji-2}
K. Yamaji, Synth. Met. \textbf{13}, 29 (1986).



\bibitem{Chaikin-1}
P.M. Chaikin, Mu-Yong Choi, J.F. Kwak, J.S. Brooks, K.P. Martin,
M.J. Naughton, E.M. Engler, and R.L. Greene, Phys. Rev. Lett.
\textbf{51}, 2333 (1983).


\bibitem{Ribault}
M. Ribault, D. Jerome, J. Tuchendler, C. Weyl, and K. Bechgaard,
J. Phys. (Paris) Lett. \textbf{44}, L-953 (1983).


\bibitem{Heritier}
M. Heritier, G. Montambaux, and P. Lederer, J. Phys. (Paris) Lett.
\textbf{45}, L-943 (1984).



\bibitem{Gorkov-2}
A.A. Abrikosov, L.P. Gor'kov, and I.E. Dzyaloshinskii, {\it
Methods of Quantum Field Theory in Statistical Mechanics} (Dover,
New York, 1963).



\bibitem{Gradshein}
I.S. Gradshteyn and I.M. Ryzhik,  {\it Table of Integrals, Series,
and Products} (6-th edition, Academic Press, London, United
Kingdom, 2000).

\bibitem{Miura-1}
A.S. Dzurak, B.E. Kane, R.G. Clark, N.E. Lumpkin, J. O'Brien, G.R.
Facer, R.P. Starrett, A. Skougarevsky, H. Nakagawa, N. Miura, Y.
Enomoto, D.G. Rickel, J.D. Goettee, L.J. Campbell, C.M. Fowler, C.
Mielke, J.C. King, W.D. Zerwekh, A.I. Bykov, O.M. Tatsenko, V.V.
Platonov, E.E. Mitchell, J. Hermann, and K.-H. Muller, Phys. Rev.
B \textbf{57}, R14084 (1998).

\bibitem{Miura-2}
T. Sekitani, N. Miura, S. Ikeda, Y.H. Matsuda, and Y. Shiohara,
Physica B \textbf{346-347}, 319 (2004).

\bibitem{Chaikin-2}
W. Kang, S.T. Hannahs, and P.M. Chaikin, Phys. Rev. Lett.
\textbf{70}, 3091 (1993).


\end{references}
\end{document}